\documentclass[aps,prl,twocolumn,superscriptaddress,10pt,floatfix,amsmath,amssymb]{revtex4-2}
\usepackage{graphicx} 
\usepackage{dcolumn}
\usepackage{xcolor}
\usepackage{bm}
\usepackage{physics}
\usepackage{lipsum}  
\usepackage{hyperref}
\usepackage[normalem]{ulem}
\usepackage{braket}
\usepackage{orcidlink}
\graphicspath{{images_rev/}}

\newcommand{\oper}[1]{\hat{\mathcal{#1}}}
\newcommand{\sect}[1]{\textit{#1}.--}

\begin{document}

\title{
Boosting Work Extraction in Quantum Batteries via Continuous Environment Monitoring
}

\date{\today}
\author{Gabriele Cenedese}
    \email{gcenedese@uninsubria.it}
    \affiliation{Center for Nonlinear and Complex Systems, Dipartimento di Scienza e Alta Tecnologia, Universit\`a degli Studi dell'Insubria, via Valleggio 11, 22100 Como, Italy} 
    \affiliation{Istituto Nazionale di Fisica Nucleare, Sezione di Milano, via Celoria 16, 20133 Milano, Italy}

\author{Giuliano Benenti}
    \email{giuliano.benenti@uninsubria.it}
    \affiliation{Center for Nonlinear and Complex Systems, Dipartimento di Scienza e Alta Tecnologia, Universit\`a degli Studi dell'Insubria, via Valleggio 11, 22100 Como, Italy} 
    \affiliation{Istituto Nazionale di Fisica Nucleare, Sezione di Milano, via Celoria 16, 20133 Milano, Italy}

\author{Dario Ferraro}
\email{dario.ferraro@unige.it}
\affiliation{Dipartimento di Fisica, Università di Genova, Via Dodecaneso 33, 16146 Genova, Italy}
\affiliation{CNR-SPIN, Via Dodecaneso 33, 16146 Genova, Italy}

\author{Marco G. Genoni\,\orcidlink{0000-0001-7270-4742}}
\email{marco.genoni@unimi.it}
\affiliation{Dipartimento di Fisica “Aldo Pontremoli”, Università degli Studi di Milano, I-20133 Milan, Italy}

\begin{abstract}
During the charging process, interactions between a quantum battery and its charger generally generate quantum correlations, which may reduce the amount of work extractable from the battery alone.
We show that, by coupling the system with an environment that can be continuously monitored, one can weaken these correlations and enhance work extraction beyond what is achievable in the ideal (closed system) limit.
This general mechanism is illustrated using both a cavity--mediated spin--spin and a Dicke quantum battery model.
\end{abstract}
\maketitle
\sect{Introduction} 
Advances in quantum science and technology for computation~\cite{benenti2019principles}, simulation~\cite{georgescu2014quantum}, sensing~\cite{degen2017sensing} and communication~\cite{portmann2022security} 
have driven growing interest in leveraging quantum--mechanical resources---such as coherence and entanglement---for energy storage and manipulation, giving rise to the concept of the Quantum Battery (QB)~\cite{alicki2013entanglement,Quach23, Campaioli24,Ferraro26, Binder15, campaioli2017enhancing, Ferraro18,Andolina18,Le18,Zhang19, Hovhannisyan20,Seah21,gyhm2022quantum,salvia2022extracting,morrone2023charging,castellano2024extended,shaghaghi2022micromasers,grazi2024controlling, catalano2024frustrating,castellano2024extended, konar2024multimode,gemme2024qutrit,razzoli2025cyclic,lu2025topological, beder2025workextractionquantumbattery,cavaliere2025dynamical, cavaliere2025quantumadvantageboundsmultipartite} 
A QB is a quantum system that can be driven from a low--energy configuration (empty battery, ideally the ground state) to an excited state (full battery)
through interaction with an external charger, and can subsequently release the stored energy on demand.

On very general grounds, when the charger is another quantum system, the total Hamiltonian of the composite system can be expressed as $\oper{H}(t) = \oper{H}_B + \oper{H}_C + \oper{H}_{\text{int}}(t)$,
where $\oper{H}_B$ describes the battery ($B$), $\oper{H}_C$ the charger ($C$), and $\oper{H}_{\text{int}}(t)$ accounts for a time--dependent interaction between them.
In the simplest charging protocol, this coupling is switched on only during the charging stage, which has a controllable duration $\tau_c$.

After that, the charger is disconnected, and the battery is left in the state 
$\rho_B(\tau_c)$, storing energy 
$E[\rho_B(\tau_c)]=\text{Tr}_B\bigl[\oper{H}_B \rho_B(\tau_c)\bigr]$. To ensure reversibility and prevent dissipative losses, the extraction 
of useful work must then be carried out through unitary operations~\cite{bhattacharjee2021quantum}. 
The \emph{ergotropy}~\cite{allahverdyan2004maximal} $\mathcal{E}$
quantifies the maximum amount of energy that can be converted into work 
via coherent, reversible dynamics:
$\mathcal{E}[\rho_B(\tau_c)] = E[\rho_B(\tau_c)]- \text{Tr}_B\bigl[\oper{H}_B {\rho}_{B,p}(\tau_c)\bigr]$, where $\rho_{B,p}(\tau_c)$ is the passive state 
unitarily connected to $\rho_B(\tau_c)$ and from which no additional work can be obtained~\cite{Lenard78, Pusz78}. 

As with all quantum devices, realistic QBs do not evolve in isolation. This observation has motivated the study of open QBs, which exchange both energy and information with their surrounding environment. 
Such an environment induces decoherence and dissipation, processes typically associated with losses of energy and ergotropy~\cite{Farina19,caravelli2021energy,zakavati2021bounds,shaghaghi2023lossy,yadav2025thermo}, although engineered reservoirs have also been proposed as a useful resource~\cite{ahmadi2025superoptimal,xu2021enhancing,song2022environment,yu2023enhancement,centrone2023charging, Ahmadi24}.

A promising way to mitigate harmful environmental impacts is through continuous monitoring of the environment~\cite{albarelli2024,wiseman2009quantum}. A wide range of theoretical studies has examined this idea for different purposes, from feedback-based schemes for quantum-state preparation and measurement~\cite{wisemanQuantumTheoryContinuous1994,dohertyFeedbackControlQuantum1999,ThomsenSpinSqueezingQuantum2002,Gambetta08, WisemanDoherty,genoniQuantumCoolingSqueezing2015,Korotkov16, brunelliConditionalDynamicsOptomechanical2019,digiovanniUnconditionalMechanicalSqueezing2021,candeloroFeedbackAssistedQuantumSearch2023,isaksenMechanicalCoolingSqueezing2023,Caprotti_2024} to quantum estimation protocols~\cite{mabuchiDynamicalIdentificationOpen1996,geremiaQuantumKalmanFiltering2003,tsangOptimalWaveformEstimation2009,gammelmarkFisherInformationQuantum2014,sixParameterEstimationMeasurements2015,kiilerichBayesianParameterEstimation2016,genoniCramErRaoBound2017,albarelliUltimateLimitsQuantum2017,Albarelli2018restoringheisenberg,rossiNoisyQuantumMetrology2020,AmorosBinefa2021,FallaniPRXQuantum2022,ilias2022criticality,yangEfficientInformationRetrieval2022,amorosbinefa2024,midha2025metrologyopenquantumsystems,khan2025tensornetworkapproachsensing,yang2025quantumcramerraoprecisionlimit}, and in non-equilibrium quantum thermodynamics~\cite{manzanoQuantumThermodynamicsContinuous2022,garrahanThermodynamicsQuantumJump2010,rossiExperimentalAssessmentEntropy2020,landiInformationalSteadyStates2022}.
Complementing these theoretical developments, experiments have realized related protocols in superconducting devices~\cite{murchObservingSingleQuantum2013,campagne-ibarcqObservingQuantumState2016,ficheuxDynamicsQubitSimultaneously2018,minevCatchReverseQuantum2019, Hacohen-Gourgy20}, optomechanical systems~\cite{wieczorekOptimalStateEstimation2015,rossiObservingVerifyingQuantum2019,rossiMeasurementbasedQuantumControl2018,magriniRealtimeOptimalQuantum2021}, and atomic settings~\cite{Fama24, duan_concurrent_2025}.

In particular, in the context of QBs, by exploiting the information extracted from measurements, it becomes possible to design feedback--based protocols that improve either the charging process~\cite{mitchison2021charging,yao2022optimal} or the work extraction~\cite{morrone2023,Elyasi25,kua2025daemonicergotropygaussianquantum,crotti2026}.

In this Letter, we address a more ambitious question: can coupling our system to a continuously monitored environment be exploited to enhance work extraction beyond what is achievable in the ideal, noiseless case?

A positive answer is suggested by the following reasoning.
Consider first a closed system composed of the QB and the charger.
During the charging process, if the interaction generates entanglement, or more generally, quantum correlations between the two subsystems, 
the work extractable from the battery alone may be reduced~\cite{Shi22, gyhm2024beneficial}. Dissipation and continuous environment monitoring
can weaken these quantum correlations and improve energy--extraction performance, as the 
information gained through the measurement process can be used to increase the extractable work.
We illustrate this mechanism through two representative examples:
(i) a minimal model consisting of two spins, one acting as the QB and the other as the charger, coupled via a lossy cavity mode~\cite{Sillanpaa07, Peng14,crescente2022enhancing, crescente2024boosting}; and
(ii) the Dicke model---one of the most studied and experimentally relevant QB architectures~\cite{Ferraro18, Ferraro19, andolina2019quantum, julia2020bounds, Dou22, Dou22b, quach2022superabsorption, hymas2025experimental, kurman2025quantum,yang2024three,wang2024deep,gemme2023off,erdman2024reinforcement,andolina2019,carrasco2022collective,canzio2025single}---with cavity--mode leakage. 

In both cases, we couple the system to a detector---either a photodetector or a homodyne detector---that continuously monitors the stream of leaked photons. Remarkably, the resulting \emph{daemonic ergotropy}~\cite{francica2017daemonic,Manzano2018,morrone2023}, namely the extractable work that incorporates the information gained through measurement, 
not only compensates for dissipative losses but, in certain parameter regimes, even surpasses the ideal, lossless value.

\sect{Continuously monitored open quantum systems} When the environment of an open quantum system is continuously monitored, the resulting dynamics becomes conditioned on the inherently stochastic measurement record. This gives rise to a stochastic evolution of the system’s wavefunction, commonly known as a quantum trajectory~\cite{wiseman2009quantum,albarelli2024}. Regardless of the specific measurement performed, averaging over all possible outcomes—and thus over all trajectories— is completely equivalent to tracing over the environmental degrees of freedom, thus restoring the original open-system evolution~\cite{wiseman2009quantum,albarelli2024}, which in our case is always described by a Markovian Lindblad master equation~\cite{breuer2002theory}. In what follows, we will refer to states evolving under continuous monitoring as conditional, and those following the unmonitored open-system dynamics as unconditional.
If our goal is to charge a quantum battery, the information obtained from continuous measurements can be used in two distinct ways: either during the charging stage, where real--time feedback is applied to steer the quantum state toward one with greater ergotropy~\cite{mitchison2021charging,yao2022optimal}, or, as we pursue in this work, during the extraction stage, where the work--extraction unitary is tailored to each conditional state according to the corresponding measurement outcomes~\cite{francica2017daemonic,Manzano2018,morrone2023,Elyasi25,crotti2026,kua2025daemonicergotropygaussianquantum}. In the latter case, we talk about daemonic ergotropy in analogy with the classical Maxwell daemon thought experiment~\cite{junior2025friendlyguideexorcisingmaxwells}. Suppose that $\rho_{B|k}$ is the conditional battery state given the measurement outcome $k$ with probability $p_k$. We assume that a daemon can perform a $k$--dependent unitary operation $\oper{U}_k$ optimized for each conditional state in order to extract the amount of work $\mathcal{W}_k(\rho_{B|k})$. The average extractable work is then 
\begin{equation}
 \overline{\mathcal{W}}\equiv \mathbb{E}[\mathcal{W}]=\sum_kp_k \mathcal{W}_k(\rho_{B|k})\leq \sum_kp_k \mathcal{E}(\rho_{B|k})=\overline{\mathcal{E}},
\end{equation}
where the upper bound is the daemonic ergotropy, given by the ensemble average of the ergotropy of the conditional states, $\overline{\mathcal{E}}\equiv \mathbb{E}[\mathcal{E}]$.
As shown in Refs.~\cite{francica2017daemonic,morrone2023}, the daemonic ergotropy is constrained by
$\mathcal{E}(\rho_B) \le \overline{\mathcal{E}}\le E(\rho_B)$, namely it cannot exceed the battery’s total energy $E(\rho_B)$ (which we set equal to zero in the ground state) and remains at least as large as the ergotropy of the unconditional state $\mathcal{E}(\rho_B)$, demonstrating how daemonic protocols allow one to extract energy even from the passive part of the state.

In general, different types of measurements lead to different forms of stochastic master equations, depending on the specific measurement operators involved. In the following we will assume the initial state of the system to be pure and a unit-efficiency measurement (see End Matter for details on the inefficient measurement case), such that the conditional evolution remains confined to the manifold of pure states and leads to the following Stochastic Schrödinger Equation (SSE). In the case of continuous photo--detection (PD) on the environment, one obtains the following SSE \cite{wiseman2009quantum, albarelli2024}, (from now on we assume $\hbar=1$):
\begin{eqnarray}
  \text{d}\ket{\Psi^{(c)}_{B,C}(t)}  &=& \biggl[-i \oper{H}(t) \text{d}t +\frac{1}{2} \left(  \langle \hat{c}^\dagger \hat{c} \rangle - \hat{c}^\dagger \hat{c} \right) \text{d}t   +\nonumber\\
  &+& \left( \frac{\hat{c}}{\sqrt{\langle \hat{c}^\dagger \hat{c} \rangle}} - 1 \right)\text{d}N_t \biggr] \ket{\Psi^{(c)}_{B,C}(t)},
\end{eqnarray}
where $\ket{\Psi^{(c)}_{B,C}(t)}$ is the conditional battery--charger state at time $t$ and the jump operator $\hat{c}$ depends on the specific system under consideration; $\text{d}N_t$ denotes a Poisson increment, which takes values $0$ (no detection) or $1$ (detection), and which is fully determined by its average value $\mathbb{E}[dN_t]= \langle \hat{c}^\dagger \hat{c} \rangle_t \,dt$ (we denote with $\langle\hat{A}\rangle_t = \bra{\Psi^{(c)}_{B,C}(t)} \hat{A} \ket{\Psi^{(c)}_{B,C}(t)}$ the average value of an operator on the conditional state at time $t$).

In the case of homodyne detection (HD), the measurement process corresponds to a continuous monitoring of an environment field quadrature, defined by the local oscillator phase~\cite{albarelli2024}, which from now on, we will fix as $\theta=0$. The corresponding SSE reads: 
\begin{eqnarray}
   \text{d}\ket{\Psi^{(c)}_{B,C}(t)} \nonumber &=& \biggl[-i \oper{H}(t)\text{d}t + \\  &-&\frac{1}{2} \left( \hat{c}^\dagger \hat{c} -  \hat{c}\langle \hat{c}+\hat{c}^\dagger  \rangle  +\frac{\langle \hat{c}+\hat{c}^\dagger  \rangle^2}{4} \right) \text{d}t + \nonumber\\
   &+& \left( \hat{c}-\frac{\langle \hat{c}+\hat{c}^\dagger  \rangle}{2}  \right) \text{d}w_t \biggr]\ket{\Psi^{(c)}_{B,C}(t)},
\end{eqnarray}
where $\text{d}w_t$ is a Wiener increment, namely a Gaussian stochastic variable with zero mean and variance $\text{d}t$ representing the intrinsic noise contribution to the measured photocurrent arising from the continuous quantum measurement, in formula $dy_t = \langle \hat{c}+\hat{c}^\dagger  \rangle_t \,dt + dw_t$.

For both detection schemes, as mentioned above, the unconditional state can be recovered by averaging over the ensemble of conditional pure state trajectories, that is, by performing the unraveling of the stochastic evolution, $\mathbb{E}\bigl[ \ket{\Psi^{(c)}_{B,C}(t)} \bra{\Psi^{(c)}_{B,C}(t)} \bigr]=\rho_{B,C}(t)$, and which obeys the Markovian master equation $d\rho_{B,C}(t)/dt = - i [\oper{H},\rho_{B,C}(t)] + \hat{c} \rho_{B,C}(t) \hat{c}^\dagger - \{\hat{c}^\dagger \hat{c},\rho_{B,C}(t)\}/2$. 
The conditional and unconditional states of the battery are obtained by performing the trace over the charger degrees of freedom, that is respectively as $\rho_B^{(c)}(t)=\text{Tr}_C\bigl[\ket{\Psi^{(c)}_{B,C}(t)} \bra{\Psi^{(c)}_{B,C}(t)} \bigr]$ and $\rho_B(t)=\text{Tr}_C\bigl[\rho_{B,C}(t) \bigr]$.

\begin{figure}[h!]
	\includegraphics[width=\columnwidth]{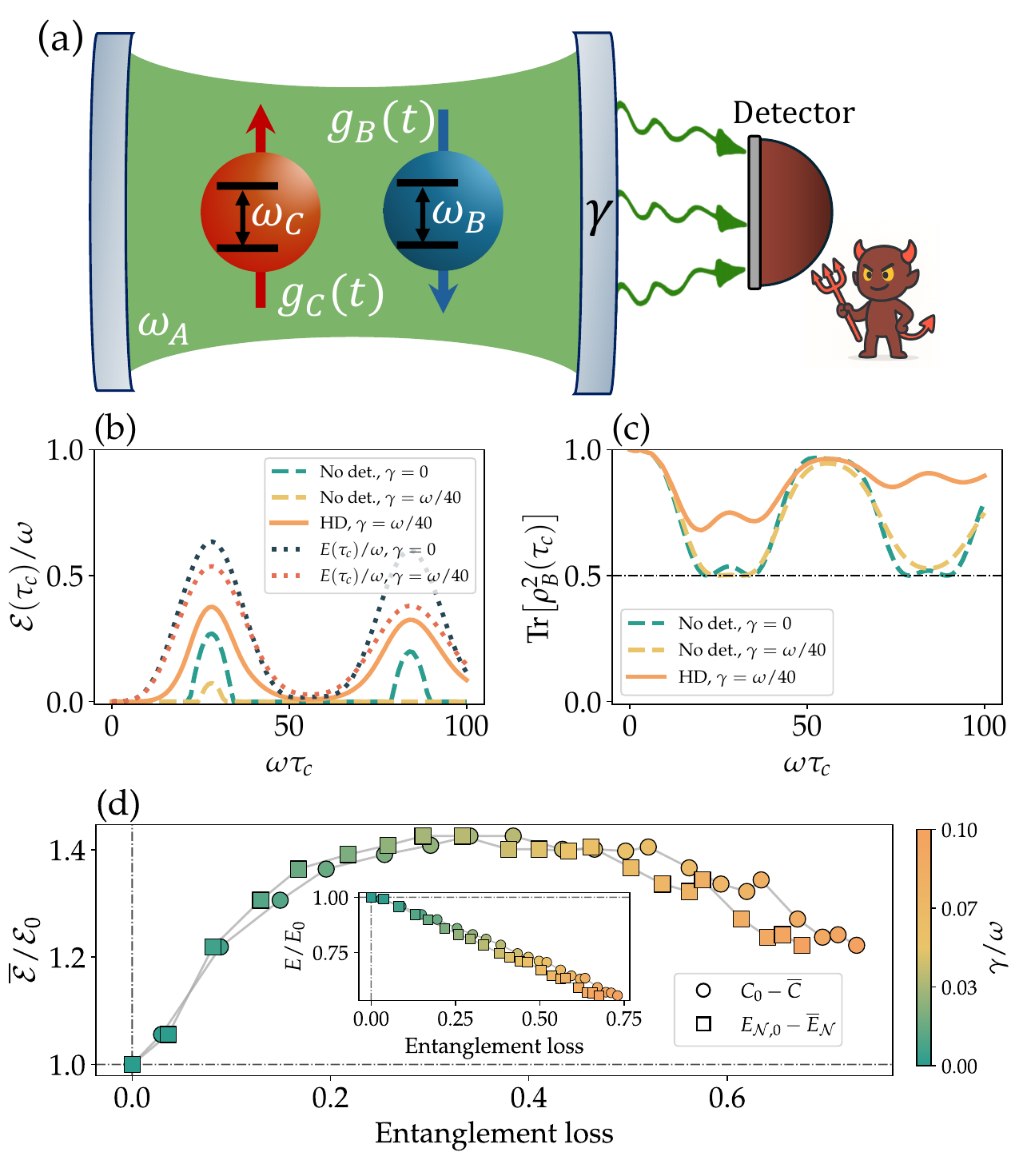}
	\caption{(a) Schematic of a cavity--mediated spin–spin QB. (b) Unconditional and daemonic ergotropies (in units of $\omega$) (dashed and solid lines) and their corresponding energy upper bounds (dotted lines). (c) Purity of the battery qubit as a function of the charging time. The black dotted--dashed line indicates the minimum achievable purity. (d) Maximum daemonic ergotropy in units of the maximum ergotropy in the noiseless case ($\mathcal{E}_{0}$) as a function of the environment-induced loss of entanglement between battery and charger quantified by means of concurrence (circles) and logarithmic negativity (squares). Inset: fraction of energy lost during the same process. Coupling strengths are $\bar{g}_B=\omega/20$, $\bar{g}_C = \omega/10$. Here and in the following figures, conditional averages
    are over $n=1000$ trajectories, unless otherwise specified.
    }
	\label{fig:1}
\end{figure} 

\sect{Results} 
We first consider a minimal model, a cavity--mediated spin--spin QB (see Fig. \ref{fig:1} (a) for a pictorial description), in which one spin acts as a battery, the other as a charger, and their interaction is mediated by a cavity mode~\cite{Sillanpaa07,Peng14,  crescente2022enhancing,  crescente2024boosting}. 
This model can be implemented, for example, using superconducting qubits coupled to a microwave resonator~\cite{Sillanpaa07}, a physical platform where continuous monitoring is nowadays a standard experimental procedure~\cite{murchObservingSingleQuantum2013,campagne-ibarcqObservingQuantumState2016,ficheuxDynamicsQubitSimultaneously2018,minevCatchReverseQuantum2019, Hacohen-Gourgy20}.
The battery, charger, and interaction Hamiltonians are given by
\begin{eqnarray}
\oper{H}_B &=& \frac{1}{2}\omega_B  \left(\hat{\sigma}_B^z+1\right), \nonumber\;\;
\oper{H}_C = \frac{1}{2}\omega_C \left(\hat{\sigma}_C^z+1\right),\\
\oper{H}_{\text{int}}(t) &=& 
\omega_A \hat{a}^\dagger \hat{a} +[g_B (t)\hat{\sigma}_B^{x} + g_C(t) \hat{\sigma}_C^{x}]( \hat{a}+ \hat{a}^\dagger),
\end{eqnarray}
where $\hat{a}$ and $\hat{a}^\dagger$ are the annihilation and creation operators of a cavity mode with frequency $\omega_A$, $\hat{\sigma}_{C(B)}^{\alpha}$ (with $\alpha=x,y,z$) are the Pauli matrices describing a charger (battery) spin with level spacing $\omega_C$ ($\omega_B$) and the cavity--mediated interaction terms are taken to be active only during the charging time $\tau_c$, when $g_B(t)=\bar{g}_B$ and $g_C(t)=\bar{g}_C$.
The cavity is initially prepared in the zero--photon Fock state $\ket{0}_A$, while the charger is excited and the battery is in the ground state:
\begin{equation}
    \ket{\psi(0)}=\ket{\downarrow}_B \otimes \ket{\uparrow}_C \otimes \ket{0}_A.
\end{equation}
From now on we consider the resonant case $\omega_C = \omega_B = \omega_A = \omega$ and we set $\overline{g}_B = \omega/20$, allowing the coupling strength 
$\overline{g}_C$ to vary. 
When $\overline{g}_C = \overline{g}_B$ the system is in perfect resonance and a complete energy transfer from $C$ towards $B$ can be achieved, while if $\overline{g}_C \neq \overline{g}_B$ the energy transfer is not optimal and correlations are generated between $B$ and $C$. Here, the dissipation is modeled as a lossy cavity with dissipation rate $\gamma$. The corresponding jump operator is $\hat{c} = \sqrt{\gamma} \hat{a}$, and we consider as an example of unraveling of this unconditional evolution the HD SSE.

The ergotropy for different values of the dissipation rate, together with the corresponding daemonic ergotropy, is shown in Fig. \ref{fig:1} (b) in the weak coupling for $\bar{g}_C=\omega/10$  as a function of the charging time (for the dependence on $\bar{g}_C$, see End Matter). Furthermore, in Fig.\ref{fig:1} (c), we can observe the purity of the battery. The battery ergotropy displays pronounced peaks that are eventually surpassed by the daemonic ergotropy.  
The interplay of dissipation and measurements leads to a purification of the battery state, thereby enhancing the amount of extractable energy. 
This enhancement is quantified in Fig. \ref{fig:1} (d), where we plot the ratio between the average conditional maximum ergotropy, $\overline{\mathcal{E}}$, and the maximum ergotropy in the noiseless case, $\mathcal{E}_0$ by increasing $\gamma$. The gain is correlated with the loss of entanglement between $B$ and $C$, quantified through the variation of two entanglement monotones: the concurrence, computed between the $B$ and $C$ qubits after tracing out the cavity, and the logarithmic negativity of the battery state. We denote by $C_0$ ($E_{\mathcal{N},0}$) the concurrence (logarithmic negativity) in the dissipationless case and by $\overline{C}$ ($\overline{E}_{\mathcal{N}}$) its average conditional value. As shown in the figure, there is a clear correlation between dissipation-induced entanglement loss and ergotropy gain, up to a threshold beyond which energy losses through dissipation become dominant.

In the PD case, the behavior is qualitatively similar, although the peak value of the daemonic ergotropy is reduced.

The enhancement of ergotropy under dissipation and 
continuous monitoring of the environment is also obtained for the Dicke QBs. The Dicke model~\cite{dicke1954coherence} 
was originally introduced to describe matter--radiation interactions across systems ranging from molecular ensembles to solid--state platforms~\cite{kirton2019introduction}, paving the way to both cavity~\cite{haroche2006exploring} and circuit quantum electrodynamics~\cite{blais2021circuit}.
It consists of an optical cavity (the charger) with a single mode of the electromagnetic field coupled in the dipole approximation to an ensemble of $N$ non--interacting Two--Level Systems (TLSs) (the battery): 
\begin{eqnarray}
\oper{H}_B &=& \omega_a \oper{S}^z +\frac{N}{2}\omega_{a},\;\;
\oper{H}_C = \omega_c \hat{a}^\dagger \hat{a},\nonumber\\
\oper{H}_{\text{int}}(t) &=& 2 \lambda(t) \oper{S}^x(\hat{a}+\hat{a}^\dagger).
\label{H_Dicke}
\end{eqnarray}
Here the cavity mode has frequency $\omega_c$, $\oper{S}^\alpha=\frac{1}{2}\sum_i^N\hat{\sigma}_i^\alpha$ (with $\alpha=x,y,z$) are the component of the collective spin operator expressed in terms of the Pauli matrices $\hat{\sigma}_i^\alpha$ of the $i$--th TLS. Being $\ket{g}$ and $\ket{e}$ the ground and the excited state of a single TLS, the quantity $\omega_a$ is the energy splitting between $\ket{g}$ and $\ket{e}$. We again consider a dissipative cavity described by the jump operator $\hat{c}=\sqrt{2\kappa}\hat{a}$.
 
The interaction strength is governed by the parameter $\lambda(t)$, whose explicit time dependence is determined by the specific charging  protocol. Here, we focus on the resonant regime $\omega_a=\omega_c=\omega$ (see End Matter for considerations about the non-resonant case).


\begin{figure}[h!]
	\includegraphics[width=\columnwidth]{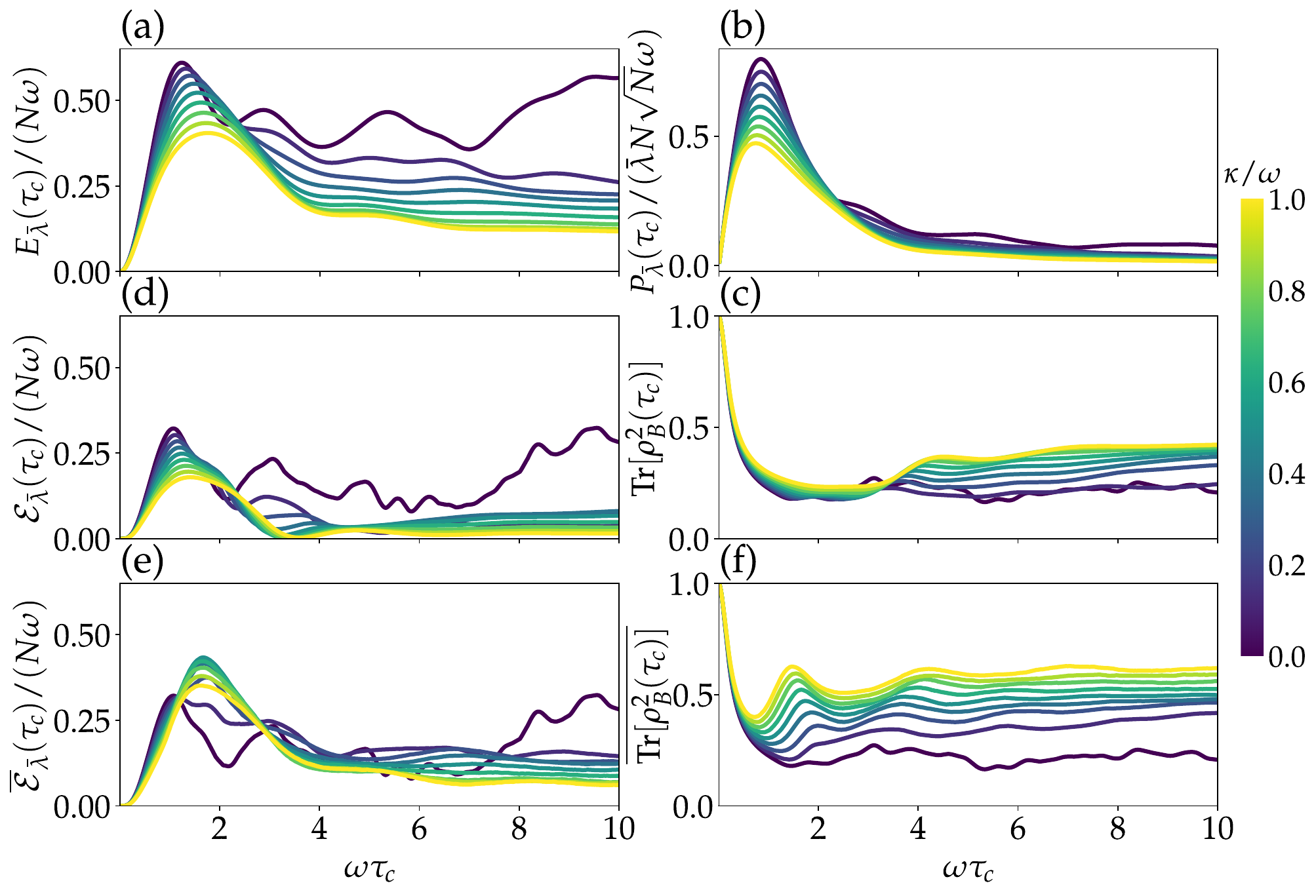}
	\caption{(a–d) Energy, power, ergotropy and purity of the reduced Dicke battery state as a function of the charging time for the unconditional dynamics. (e,f) Daemonic ergotropy and purity for the conditional dynamics under the PD scheme. Parameter values: $\bar{\lambda} = 0.3\omega $, $ N = 6 $. Note that energy, ergotropy and power are rescaled to make them adimensional and highlight the well--known scalings laws~\cite{Ferraro18}
$E_{\bar{\lambda}},\mathcal{E}_{\bar{\lambda}}\propto N$ and 
    $P_{\bar{\lambda}}\propto N^{3/2}$. 
    }
	\label{fig:2}
\end{figure} 

As initial state of the system we assume 
\begin{equation}
    \ket{\Psi^{(N)}(0)}=\ket{N}\otimes\ket{g}^{\otimes N},
\end{equation}
where the single cavity mode is initialized in the $N$--photon Fock state $\ket{N}$ while the battery state is the tensor product of the ground states of the individual TLSs.
We consider the simplest charging protocol, which consists of suddenly switching on the interaction at time $t=0$ and keeping it active until charging time $\tau_c$. The interaction strength is fixed at a constant value $\bar{\lambda}$ during charging.
Since the number of photons in the cavity is neither conserved nor inherently bounded due to the presence of counter--rotating terms in the Hamiltonian in Eq.~(\ref{H_Dicke}), we need to introduce a cutoff in the dimension of the Fock Hilbert space by limiting the maximum photon number. In the following, we set $N_{ph}=20$ for $N<5$, and $N_{ph}=4N$ for $N\geq5$. We checked that with this choice numerical convergence is ensured. 

\begin{figure}[h!]	\includegraphics[width=\columnwidth]{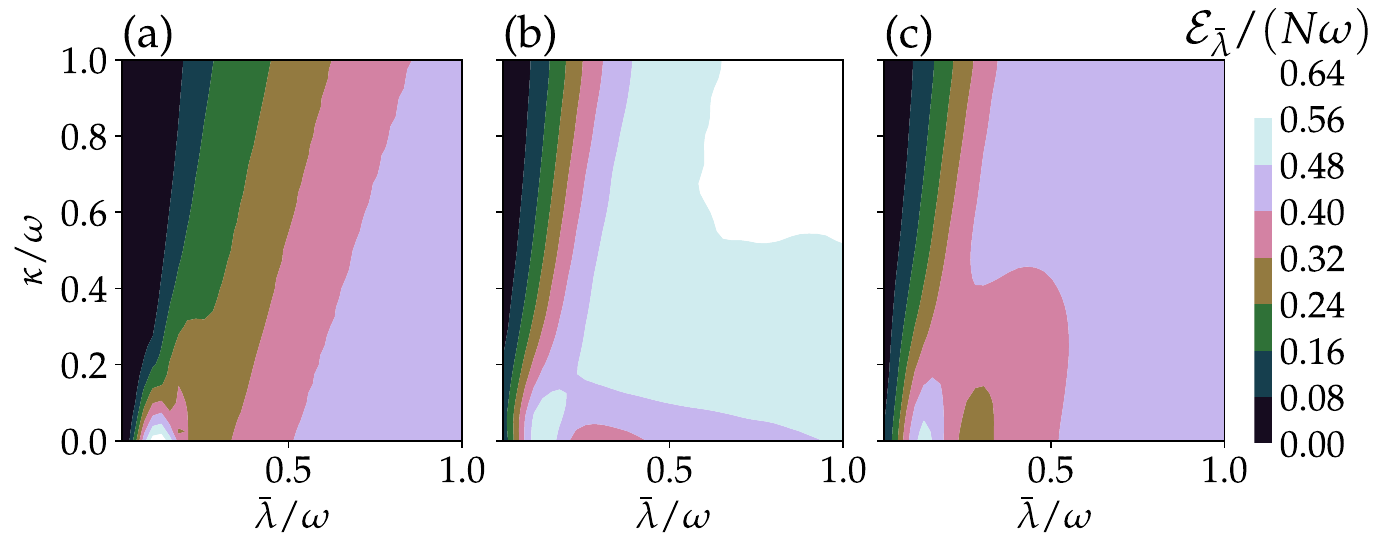}
	\caption{Contour plots of ergotropy in the unmonitored scenario (a), HD daemonic ergotropy (b), and PD daemonic ergotropy (c) as a function of the coupling strength $\bar{\lambda}$ and the dissipation rate $\kappa$, for $N=6$. Note that ergotropies are calculated at the charging times for which the stored energy is at its maximum.}
	\label{fig:3}
\end{figure} 

First, we computed the battery energy $E_{\bar{\lambda}}(\tau_c)$, 
the charging power $P_{\bar{\lambda}}(\tau_c) =E_{\bar{\lambda}}(\tau_c)/\tau_c$,
the ergotropy
$\mathcal{E}_{\bar{\lambda}}(\tau_c)$ associated to the density matrix $\rho_B(\tau_c)=\text{Tr}_C\bigl[ \rho^{(N)}_{\bar{\lambda}}(\tau_c)  \bigr]$, where $\rho^{(N)}_{\bar{\lambda}}(\tau_c)$ denotes the state of the system at the end of the charging stage, and its purity as a function of charging time, for different values of 
the leakage rate $\kappa$ of the cavity mode (Fig. \ref{fig:2} (a-d)). In all cases, the maximum ergotropy is reached at an intermediate time, later than the peak of the power but earlier than the maximum of the energy. The purity of the battery state in unconditional dynamics remains generally low, especially in the low dissipation regime, which suggests the presence of strong correlations between the battery and the charger; for larger $\kappa$, the loss of purity can instead be attributed to decoherence at the global level. When considering conditional dynamics (Fig. \ref{fig:2} (e,f)), we find that the conditional purity and ergotropy are systematically enhanced with respect to the unconditional case, and in particular, for the selected parameters, the peak of the daemonic ergotropy surpasses the noiseless value for all values of $\kappa$.
 
\begin{figure}[h!]	\includegraphics[width=\columnwidth]{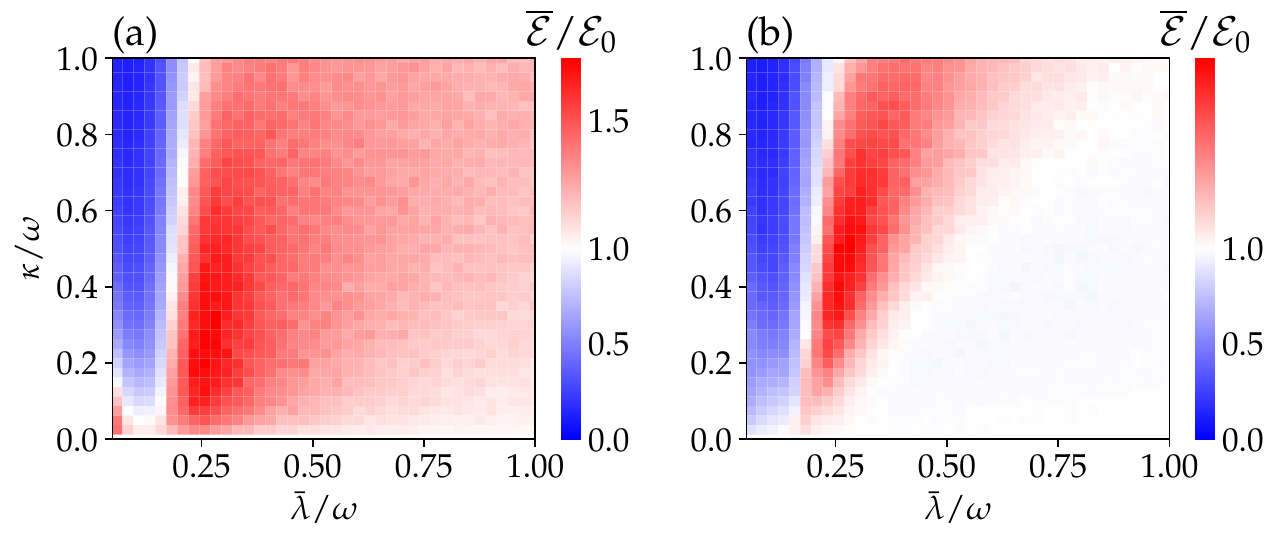}
	\caption{Daemonic ergotropy enhancement ratio as a function of the coupling strength $\bar{\lambda}$ and the dissipation rate $\kappa$ for HD (a) and PD (b),
    for $N=6$. In the red regions of the plots, the presence of dissipation combined with continuous measurement leads to an increase in daemonic ergotropy compared to the ideal dissipation--free case.} 
	\label{fig:4}
\end{figure}

Contour plots of standard and daemonic (HD and PD) ergotropies in the 
$(\bar{\lambda},\kappa)$--plane are shown in
Fig. \ref{fig:3}. 
In both HD and PD, we observe a clear enhancement of the daemonic ergotropy compared to the unconditional one. Notably, the daemonic ergotropy can also 
be higher than the ergotropy of the ideal, dissipationless case. 
In order to quantify such enhancement, we compute the ratio between the daemonic ergotropy $\overline{\mathcal{E}}$ and the ergotropy of the unconditional state without dissipation $\mathcal{E}_0$ (see Fig. \ref{fig:4}).
This comparison reveals that, over a wide region of parameter space, the presence of dissipation combined with continuous measurements leads to an increase in ergotropy compared to the ideal case without dissipation, for both the HD and PD schemes. The increase occurs in the superradiant region, where strong correlations are established between the cavity and the TLS. These correlations, which greatly limit work extraction, are reduced by monitoring the environment and, consequently, ergotropy increases. 
As shown in the End Matter, the enhancement of $\overline{\mathcal{E}}$ over $\mathcal{E}_0$ persists when varying the number $N$ of TLSs and remains robust also in the presence of TLS--cavity detuning and inefficient detection.

\sect{Conclusions}
We have shown that continuous measurements can increase the extractable work for open quantum batteries by mitigating the detrimental effects of quantum correlations between the battery and the charger and dissipation. We demonstrated that the information extracted from the environment can be effectively exploited to recover the ergotropy lost in dissipative settings and, in certain regimes, even surpass the noiseless value.
This general mechanism has been illustrated in two models of growing complexity: 
a minimal cavity--mediated spin--spin quantum battery and the Dicke quantum battery. 

More generally, our findings suggest that appropriately designed measurement protocols, tailored to the specific dissipation mechanisms of a given experimental platform, can be used as an active resource to enhance the ergotropy of quantum batteries in situations where correlations limit the extractable work. 

In the present work, the monitoring strategy is exploited at the extraction stage: the measurement record provides information about the conditional state of the battery, allowing the extraction unitary to be optimized and thereby increasing the daemonic ergotropy. 
This use of monitoring is distinct from measurement-based charging protocols, where monitoring and feedback are employed to improve the charging stage, and from stabilization protocols, where monitoring may be used to preserve a charged state against decoherence or dissipation~\cite{gherardini2020stabilizing}. 
Optimizing detection strategies for these different tasks represents an interesting direction for future work, although the optimal measurement scheme need not be the same in each case.


Finally, we comment on the thermodynamic costs associated with the protocol we have here proposed. While a complete energetic balance should ideally account for the detector’s cost and memory erasure, these factors are typically excluded from the standard definition of daemonic ergotropy used here. We note that while several attempts to quantify the cost of discrete measurements in quantum thermodynamics protocols have been presented in the literature~\cite{sagawaSecondLawThermodynamics2008,jacobsQuantumMeasurementFirst2012b,deffnerQuantumWorkThermodynamic2016b,Cottet2017,abdelkhalekFundamentalEnergyCost2018b,minagawaUniversalValiditySecond2024,Satriani_2024,latuneThermodynamicallyConsistentApproach2025,Kirchberg2025,Hagman_2025}, extending these frameworks to the continuous monitoring regime remains a non-trivial challenge. Furthermore, these energetic bounds often scale with the temperature of the environment that interacts with the detector. In platforms such circuit-QED, where a zero-temperature approximation is physically well-justified for the dynamics, such models would formally imply a vanishing measurement cost; this, however, clearly does not account for the significant practical price of maintaining the cryogenic environment. Our focus is therefore to isolate the fundamental mechanism by which continuous monitoring enhances work extraction by reducing correlations. A detailed accounting of the platform-specific trade-off between this information gain and the total refrigeration and monitoring overhead remains a promising direction for future research~\cite{Fellous-Asiani23, Meier25}.

On a broader perspective, continuous monitoring of the environment emerges not only as a tool to counteract decoherence, but also as a means to suppress unwanted correlations that naturally build up during
quantum protocols, for instance in distributed 
quantum computing architectures between
quantum processing units and quantum interconnects~\cite{Axline2018,Bienfait2019,AghaeeRad2025}.

\begin{acknowledgments}
\sect{Acknowledgments}
G. C. thanks L. Razzoli and A. Pozzoli for insightful discussions. G. B and D. F. acknowledge support from the European Union-NextGenerationEU through the ”Solid State Quantum Batteries: Characterization and Optimization” (SoS-QuBa) project (Prot. 2022XK5CPX), in the framework of the PRIN 2022 initiative of the Italian Ministry of Uni versity (MUR) for the National Research Program (PNR). This project has been funded within the programme “PNRR Missione 4- Componente 2 Investimento 1.1 Fondo per il Programma Nazionale di Ricerca e Progetti di Rilevante Interesse Nazionale (PRIN)”. M. G. G. acknowledges support from MUR and Next Generation EU via the PRIN 2022 Project CONTRABASS (Contract N.2022KB2JJM), NQSTI-Spoke2-BaC project QMORE (contract n. PE00000023-QMORE), NQSTI-Spoke1-BaC project QSynKrono (contract n. PE00000023-QuSynKrono) and NQSTI-Spoke1-BaC project QBETTER (contract n. PE0000023-QBETTER). G. B. and G.C. acknowledges support from INFN through the project "QUANTUM". 
We have numerically simulated the systems using QuTiP~\cite{johansson2012qutip}.
\end{acknowledgments}

\bibliography{Bibliography.bib}
\section{End matter}
\sect{Daemonic efficiency}
In the context of continuously monitored quantum systems, 
it is useful to introduce as additional figure of merit the daemonic efficiency, defined as 
\begin{equation}
    \eta = \frac{\overline{\mathcal{E}} - \mathcal{E}}{E - \mathcal{E}} ,
\end{equation}
where $\overline{\mathcal{E}}$ denotes the maximum daemonic ergotropy of the conditional state under consideration, while $\mathcal{E}$ represents the ergotropy of the corresponding unconditional state. The efficiency $\eta \in [0,1]$ quantifies the relative enhancement of extractable work due to measurement: it reaches unity when all the available energy is converted into extractable energy in the conditional case, and vanishes when no measurement-induced advantage is present.
\begin{figure}[h!]
    \includegraphics[width=0.85\columnwidth]{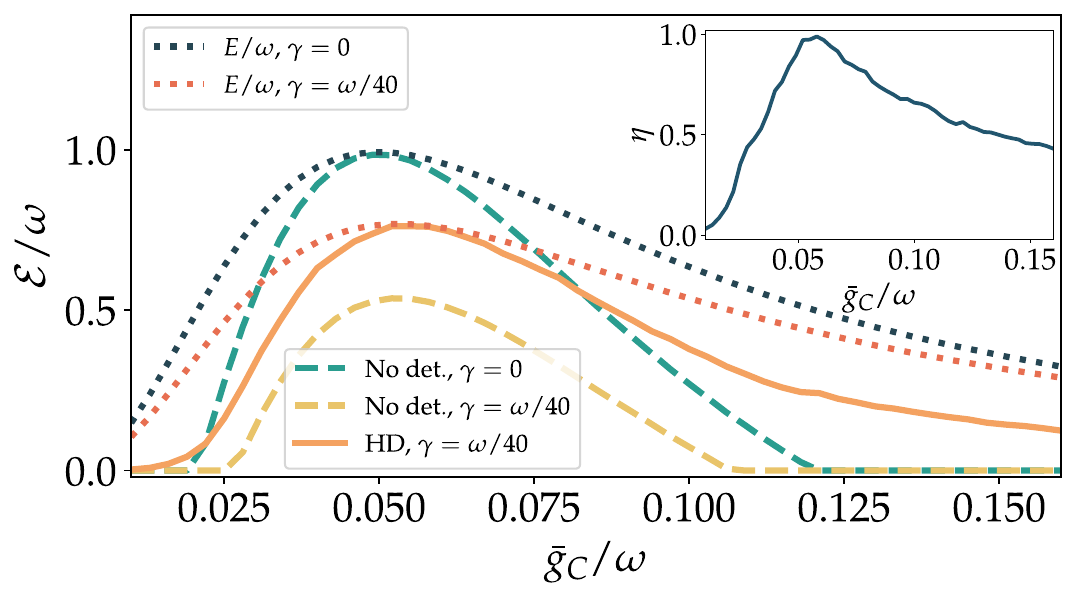}
	\caption{Maximum ergotropy as a function of $\bar{g}_C$ in the cavity--mediated spin--spin QB. The blue and red dotted lines indicate the upper energy bounds for the non-dissipative and dissipative ergotropies, respectively. 
    In the inset we show the maximum
    daemonic efficiency $\eta$ as a function of $\bar{g}_C$. Here the battery coupling strength is $\bar{g}_B=\omega/20$.}
    \label{fig:spin-cavity_ergotropy}
\end{figure}
\begin{figure}[h!]
    \includegraphics[width=0.85\columnwidth]{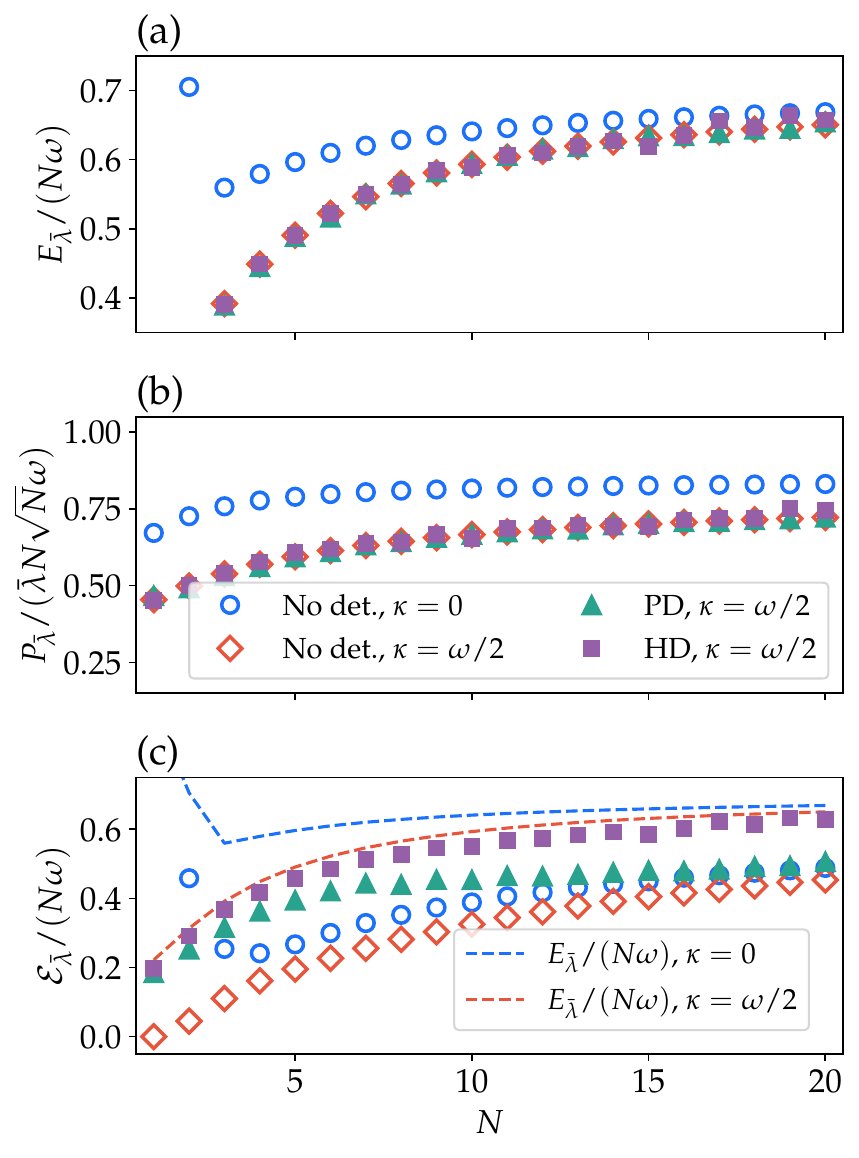}
	\caption{Maximum charging energy (a), power (b) and ergotropy (c) of the battery reduced state as functions of the system size $N$ for the Dicke QB. Open markers represent the unconditional values, while full markers correspond to the conditional values. The dashed lines in (c) represent the enery upper bounds on the ergotropy.}%
    \label{fig:scaling_conditional}
\end{figure}
\begin{figure}[h!]
    \includegraphics[width=\columnwidth]{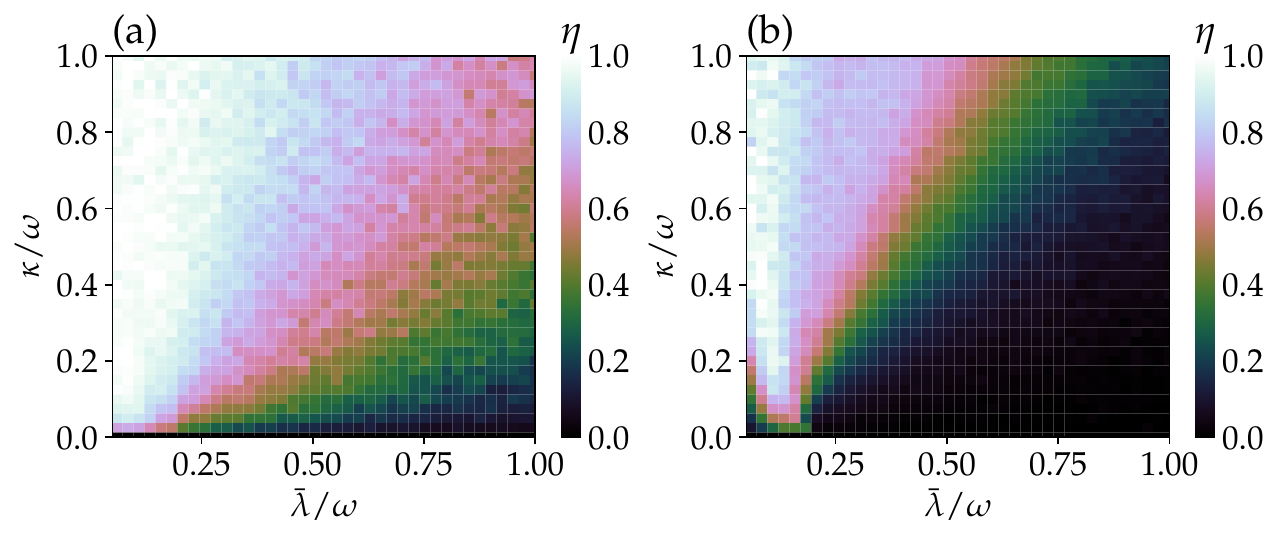}
	\caption{Daemonic efficiency of the Dicke QB as a function of the coupling strength $\bar{\lambda}$ and the dissipation rate $\kappa$ for HD (a) and PD (b), for $N=6$. In the white regions of the plots, almost all the available energy is converted into extractable daemonic ergotropy.}
    \label{fig:efficiency}
\end{figure}
\sect{Spin--spin quantum battery}
For this model, Fig.~\ref{fig:spin-cavity_ergotropy} shows the maximum ergotropy, $\mathcal{E} = \max_{\tau_c} \mathcal{E}(\tau_c)$, as a function of $\bar{g}_C / \omega \in [0.01, 0.16]$: the noiseless (unconditional) ergotropy peaks at $\bar{g}_C /\omega  \approx 0.05$, while the daemonic ergotropy surpasses it for $\bar{g}_C / \omega \gtrsim 0.1$.
The daemonic efficiency, shown in the inset of Fig.~\ref{fig:spin-cavity_ergotropy}, is generally large, indicating that daemonic feedback allows one to convert a substantial portion of the passive energy $E - \mathcal{E}$ into useful work.

\sect{Dicke quantum battery}
The scaling behavior of the Dicke QB in the conditional case is illustrated in Fig. \ref{fig:scaling_conditional}. Energy and power closely follow the trends observed in the unconditional dynamics, as expected when averaging for a sufficiently large number of trajectories, since these quantities are the same regardless of whether evaluated on the conditional or unconditional state. However, regarding the ergotropy, a striking difference emerges: for both PD and HD, the conditional ergotropy surpasses the closed--system value, with HD yielding a larger enhancement, in the same fashion as the cavity--mediated spin--spin QB.
\begin{figure}[h!]
    \includegraphics[width=0.85\columnwidth]{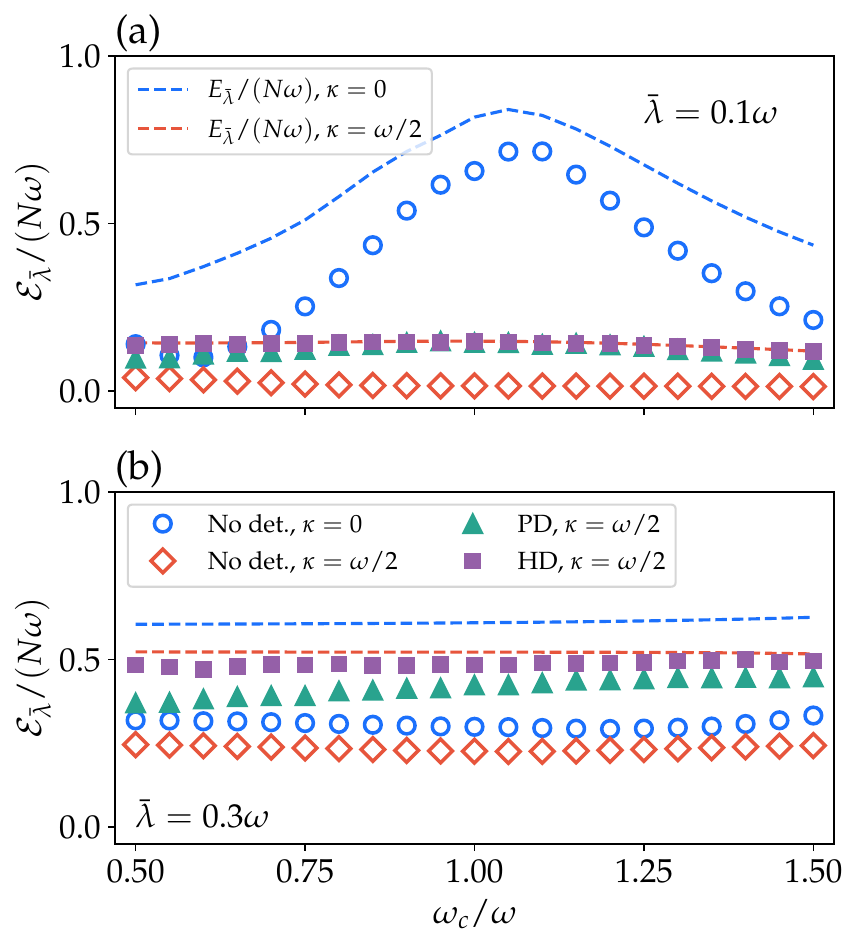}
	\caption{Maximum charging and ergotropy of the battery reduced state as functions of the detuning $\omega_c/\omega$ for the Dicke QB for $\bar{\lambda}=0.1\omega$ (a) and $\bar{\lambda}=0.3\omega$ (b). Open markers represent the unconditional values, while full markers correspond to the conditional values. The dashed lines represent the energy upper bounds on the ergotropy.}
    \label{fig:detuning_erg}
\end{figure}
\begin{figure}[h!]
    \includegraphics[width=0.85\columnwidth]{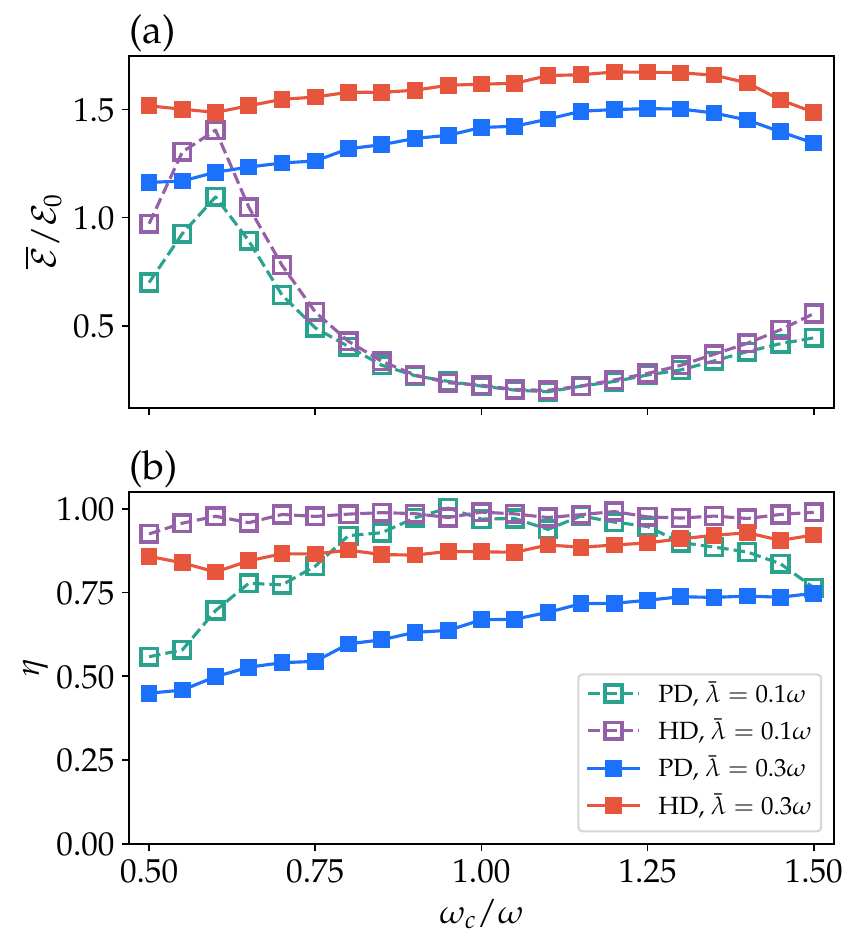}
	\caption{Daemonic ergotropy enhancement ratio (a) and daemonic efficiency (b) as a function of the cavity detuning $\omega_c/\omega$ for the Dicke quantum battery, with $\omega_a=\omega$ fixed. Here, $N=6$ and $\kappa=\omega/2$, and conditional values are averaged over $2000$ trajectories.}
    \label{fig:detuning_ratio}
\end{figure}
\begin{figure}[h!]
    \includegraphics[width=0.85\columnwidth]{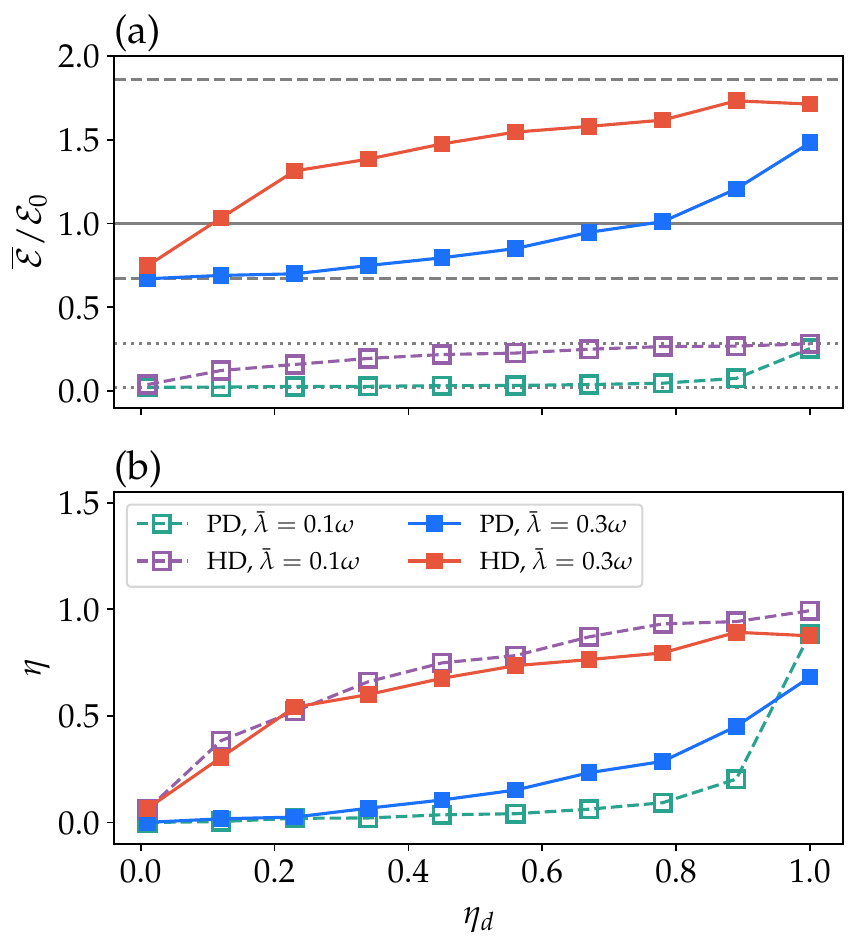}
	\caption{Daemonic ergotropy enhancement ratio (a) and daemonic efficiency (b) as a function of the detector efficiency $\eta_d$ for the Dicke quantum battery. Here, $N=4$ and $\kappa=\omega/2$, and conditional values are averaged over $200$ trajectories. The dashed lines in panel (a) correspond to the bounds of daemonic ergotropy.}
    \label{fig:ineff_detection}
\end{figure}

In Fig.~\ref{fig:efficiency}, we see that for both HD and PD the daemonic ergotropy nearly saturates the maximum extractable energy over a wide region of the parameter space (approximately corresponding to the blue regions in Fig.~\ref{fig:4}), which represents the high--dissipation regime. This does not, by itself, guarantee a large absolute value of ergotropy, but it does indicate that in this regime almost all the available energy is, in principle, fully convertible into useful work.

To test the robustness of the enhancement in the energy extraction, we first introduce cavity detuning by varying $\omega_c$ while keeping $\omega_a=\omega$ fixed. In Fig.~\ref{fig:detuning_erg} we report the results for two coupling strengths, $\bar{\lambda}=0.1\omega$ and $\bar{\lambda}=0.3\omega$, corresponding respectively to the blue and red regions in Fig.~\ref{fig:4}, with $\kappa=\omega/2$. The daemonic ergotropy enhancement ratio and the daemonic efficiency, shown in Fig.~\ref{fig:detuning_ratio}, demonstrate that the discussed effect remains robust over a broad range of detunings.

Finally, Fig.~\ref{fig:ineff_detection} shows the daemonic ergotropy enhancement ratio and the daemonic efficiency for situations where the detection efficiency is not perfect. We observe that the enhancement remains robust even for inefficient detection, with homodyne detection consistently outperforming photodetection.

\end{document}